\documentclass[preprint]{aastex}

\newcommand{\be}{\begin{equation}}
\newcommand{\ee}{\end{equation}}

\begin{document}
\title{The GJ 876 Planetary System -- A Progress Report} 

\bigskip
\author{
Gregory Laughlin\altaffilmark{5},
R. Paul Butler\altaffilmark{4},
Debra A. Fischer\altaffilmark{3},
Geoffrey W. Marcy\altaffilmark{2},
Steven S. Vogt\altaffilmark{5},
Aaron S. Wolf\altaffilmark{5}}

\email{laughlin@ucolick.org}

\altaffiltext{1}{Based on observations obtained at Lick Observatory, which
is operated by the University of California, and on observations obtained
at the W.M. Keck Observatory, which is operated jointly by the
University of California and the California Institute of Technology.}

\altaffiltext{2}{ Department of Astronomy, University of California,
Berkeley, CA USA 94720}

\altaffiltext{3}{ Department of Physics and Astronomy, San Francisco
State University, San Francisco, CA USA 94132}

\altaffiltext{4}{ Department of Terrestrial Magnetism, Carnegie Institution
of Washington, 5241 Broad Branch Road NW, Washington DC, USA 20015-1305}

\altaffiltext{5}{ UCO/Lick Observatory, University of California at Santa Cruz,
Santa Cruz, CA, USA 95064}

\begin{abstract} 

We present an updated analysis of the GJ 876 planetary system based on 
an augmented data set that incorporates 65 new high-precision radial velocities obtained
with the Keck telescope from 2001 to 2004. These new radial velocities permit
a more accurate characterization of the 
planet-planet interactions exhibited by the system. Self-consistent
three-body orbital fits (which incorporate both the estimated instrumental uncertainties
and Gaussian stellar jitter with $\sigma=6\,{\rm m\,s^{-1}}$)
continue to show that GJ 876 ``b'' (the outer planet of the system), 
and GJ 876 ``c'' (the inner planet of the system)
are participating in a stable and symmetric 2:1 resonance condition in
which the lowest order, eccentricity type mean-motion resonance variables,
$\theta_1=\lambda_c-2\lambda_b+\varpi_c$
and $\theta_2=\lambda_c-2\lambda_b+\varpi_b$ both librate around $0^{\circ}$, with amplitudes
${\vert \theta_1 \vert}_{max} = 7.0 \pm 1.8^{\circ}$, and ${\vert \theta_2 \vert}_{max} = 34 \pm 12^{\circ} $.
($\lambda_b$ and $\lambda_c$ are the mean longitudes, and $\varpi_b$ and $\varpi_c$ are the 
longitudes of periastron).
The planets are also locked in a secular resonance which causes them to librate
about apsidal alignment with ${\vert \varpi_1 - \varpi_2 \vert}_{max}=34\pm 12^{\circ}$.
The joint line of apsides for the system is precessing at a rate $\dot{\varpi}\sim-41^{\circ}\,
{\rm yr}^{-1}$.
The small libration widths of all three resonances likely point to a 
dissipative history of differential migration 
for the two planets in the system. Three-body fits to the radial velocity data set, 
combined with a Monte-Carlo analysis of synthetic data sets,
indicate that the (assumed) co-planar inclination, $i_s$, of the system
is $i_s>20^{\circ}$. Configurations
with modest mutual inclination are, however, also consistent
with the current radial velocity data.
For non-coplanar configurations, the line of nodes of the inner planet
precesses at rates of order $-4 ^{\circ} {\rm yr}^{-1}$, and in these
cases, the inner planet can be observed to transit the parent star when
either the ascending or descending node precesses through the line of sight. 
Therefore, GJ 876 ``c'' may possibly be observed to 
transit in the relatively near future even
if it is not transiting at the present time. We comment briefly on the orbital
stability of as-yet-undetected terrestrial planets in habitable orbits, and
assess the suitability of the system as a potential target for upcoming space
missions such as the Terrestrial Planet Finder.

\end{abstract}


\section{Introduction} 

GJ 876 (HIP 113020) is the lowest mass star currently known to harbor planets,
and it is accompanied by perhaps the
most remarkable exoplanetary system discovered to date.
In 1998, Marcy et al. (1998) and Delfosse et al. (1998) announced the discovery of a 
$P\sim60 \,{\rm d}$
companion orbiting the star. This planet, designated GJ 876 ``b''
is a super-Jovian object, with $M\sin(i)=1.9 M_{\rm JUP}$, and it induces a large
($K\sim210 \, {\rm m \,s^{-1}}$) radial velocity variation in its red dwarf companion.
After continued Doppler monitoring of GJ 876,
Marcy et al. (2001) announced the discovery of a second 
($M\sin(i)=0.6 \,M_{\rm JUP}$) planet
in the system. This object (designated GJ 876 ``c'') has 
a $P\sim30 \,{\rm d}$ orbit, and was identified 
to be participating in a 2:1 mean-motion resonance with the outer planet ``b''.
 
The red dwarf GJ 876 (RA=22 53, DEC=14 16) is observable from both hemispheres, 
and is distinguished by being the
fortieth-nearest stellar system, with a Hipparcos-determined distance
of 4.69 pc (Perryman et al. 1997). Its spectral type is M4 V.
Using the bolometric correction of Delfosse et al. (1998), the Hipparcos-estimated
parallax indicates a stellar luminosity of $0.0124 L_{\odot}$. The
red-dwarf mass-luminosity relation of Henry \& McCarthy (1993) therefore 
implies a mass of
$0.32 \, M_{\odot}$, and an estimated radius of $R_{\star}=0.3 \, R_{\odot}$.

A definitive identification of the resonance conditions obeyed by the planets is made possible
by the large dynamic range of the GJ 876 radial velocities.
Among the 113 Doppler velocities obtained with the Keck telescope, 
the individually estimated instrumental errors
have an average precision of $4.65 \, {\rm m \, s^{-1}}$, with individual
precision estimates ranging as low as $2.3  \, {\rm m \, s^{-1}}$.
The two planets induce velocity swings in the star of nearly $0.5 \, {\rm km \, s^{-1}}$,
and thus allow us to take full advantage of the fine Doppler precision.
Furthermore, the outer planet has been observed for more than forty orbital
periods. These fortuitous circumstances 
allow the planet-planet interactions to be probed with a
degree of refinement that is exceeded only for the planets in 
the solar system (e.g. Laplace 1799-1802) and by the planets
orbiting the 6.2 ms radio pulsar PSR B 1257+12 
(Wolszczan \& Frail 1992, Konacki \& Wolszczan 2003).

The gravitational perturbations exerted by the planets on each other
induce a significant non-Keplerian component to the orbital motion. In 
particular, the periastra of the planets precess at a rate $\dot{\varpi} \sim -41
^{\circ} {\rm yr}^{-1}$. The non-Keplerian aspects of the motion lead to a relatively
high best-fit value (currently $\sqrt{\chi^2}=2.81$) for models that use dual-Keplerian
fitting functions to model the observed radial velocity variation.
The strong planet-planet perturbations do, however, enable the construction of
dynamical fits to the radial velocity data that both improve the
$\sqrt{\chi^2}$ statistic of the orbital fit, and place the planets 
into the secular apsidal alignment resonance and deeply within the two
co-planar 2:1 mean-motion resonances (Laughlin \&
Chambers 2001, Rivera \& Lissauer 2001, Nauenberg 2002). 
The existence of this multiply resonant configuration can be understood as the
consequence of differential migration of the two planets within GJ 876's 
protoplanetary disk
(e.g. Lee \& Peale 2001, 2002; Lee 2004), and the 
presence of strong mutual interactions
between the two planets leads to a partial removal of the so-called 
$\sin (i)$ degeneracy. System configurations in which the planetary
orbits are inclined by less than $30^{\circ}$ to the plane of the sky exhibit
significantly worse fits to the radial velocity data. 

The importance of the GJ 876 system
arises because it can provide interesting constraints
for theories of planetary formation. Because the current resonant state is 
sensitive to details of the system's history, it is worthwhile to 
evaluate the degree of confidence that can be placed in the present best-fit
orbital parameters.
The plan of this short paper is thus as follows: In \S 2, we describe co-planar fits to
the radial velocity data set. These fits allow us to 
construct a detailed model of the system, and equally importantly, 
allow us to evaluate the confidence for which 
we can determine the orbital parameters.
In \S 3, we broaden our analysis to include the 
possibility of system configurations
in which the planets do not orbit in the same plane. 
In this case, the orbital angular 
momentum vectors of the planets can precess, 
and the planets can be potentially observed to 
evolve through transiting configurations.
In \S 4, we briefly discuss how our results bear on current theoretical studies of the nascent
GJ 876 planetary system.

\section{Co-planar configurations of the Two Planets}

We first assume that the planets GJ 876 ``b'' and ``c'' are in a co-planar 
configuration perpendicular to the plane of the sky ($i_s=i_b=i_c=90^{\circ}$),
and obtain self-consistent three-body
fits to the combined Lick-Keck radial velocity data set This set includes the 16 Lick
velocities listed in Marcy et al. (2001), and the 113 Keck velocities 
listed in Table 1.  All of our orbital fits are obtained using a
Levenberg-Marquardt multi-parameter minimization algorithm (Press et al. 1992) 
driving a three-body
integrator as described in Laughlin \& Chambers (2001). The best edge-on coplanar fit is
listed in Tables 2 and 3. This reference fit has twelve free parameters, including
the planetary periods, $P_b$ and $P_c$, the mean anomalies,
$M_b$ and $M_c$ at epoch JD 2449679.6316, the orbital eccentricities, $e_b$ and 
$e_c$, the longitudes of periapse $\varpi_b$ and $\varpi_c$, the planetary masses,
$m_b$ and $m_c$, and two velocity offsets, $o_1$, and $o_2$. The quantity $o_1$ is 
an offset velocity added to the first GJ 876 radial velocity, ${v_L}_1$, 
obtained with the Lick 3-meter telescope 
($t={\rm JD} 2449679.6316$,\,${v_L}_1=58.07 \,{\rm m\,s^{-1}}$; 
see Marcy et al. 2001).
The parameter $o_2$ is an offset velocity added to all of the radial velocities taken
with the Lick Telescope. It accommodates the different
velocity zero-points of the Lick and Keck telescopes.
The mass of the star is fixed at 0.32 $M_{\odot}$.
The mean longitudes, $\lambda_i$ (used to compute the mean-motion 
resonant arguments) are related to the mean anomalies and longitudes of
periapse through $\lambda_i=
\varpi_i+M_i$.

The orbital elements listed in Table 2 
are osculating orbital elements at epoch JD 2449679.6316 (the epoch of the
first radial velocity point taken at Lick Observatory in 1994.9
listed by Marcy et al. 2001) and are 
expressed in Jacobi coordinates. As explained in Rivera \& Lissauer 2001,
and Lee \& Peale 2003, Jacobi coordinates are the most natural system
for expressing multiple-planet fits to radial velocity data.
For reference, in Table 3, we
express the orbital configuration of the system (again at JD 2449679.6316)
in Cartesian coordinates. In the Cartesian system, the line of sight from
the Earth to the Star is in the negative $y$-direction, and the $y$-component
of velocity for the star relative to the system center of mass is measured,
by convention, as a negative radial velocity.

The uncertainties in the orbital fit are estimated using Monte Carlo simulation
of synthetic data sets (Press et al. 1992). In 
this procedure, we assume that the 
true configuration of the system is that given by the orbital parameters listed in
Tables 2 and 3 (i.e. the best-fit co-planar, edge-on system).
We produce 100 synthetic data sets by integrating this
assumed planetary configuration 
forward in time, sampling the stellar reflex velocity
at all of the observed epochs, and adding (in quadrature)
noise drawn from Gaussian distributions corresponding to 
(1) an assumed $\sigma=6 \, {\rm m \, s^{-1}}$
stellar jitter and (2) the individual velocity errors. Chromospherically
quiet G and K dwarfs in the ongoing radial velocity surveys typically show
RMS scatters $\sigma\sim3-5 {\rm m\,s^{-1}}$ arising from stellar jitter
(Saar, Butler, \& Marcy 1998). Nauenberg (2002) argued that excess scatter in
the dynamical fits to the GJ 876 radial velocity data should be attributed
to a stellar jitter of $2-4 {\rm m \, s}^{-1}$. Preliminary work by Wright \& Marcy (2004)
indicates that M3-M4 dwarfs with chromospheric activity similar to GJ 876 
display typical jitter values  $\sigma=4 \pm 2 \, {\rm m \, s^{-1}}$,
motivating our conservative 
choice of $\sigma=6 \, {\rm m\,s^{-1}}$.

The assumption of a $6 \,{\rm m\, s^{-1}}$
stellar jitter gives an average $\sqrt{\chi^{2}}=1.51 \pm 0.09$ for the
fits to the synthetic data sets, consistent with the value
$\sqrt{\chi^2}=1.535$ obtained from fitting to the actual data set.
(All the $\sqrt{\chi^{2}}$ values that we quote are computed using
only the instrumental uncertainties, and do not include the scatter 
expected to arise from stellar jitter). 
The uncertainty quoted for each orbital parameter is the
variance computed for the parameter from the fits to the 100 synthetic
data sets. We find that the distributions of parameter estimates
are generally consistent with underlying Gaussian distributions. We
note, however, that significant co-variation does exist between some
of the orbital parameters (e.g. $M_i$ and $\varpi_i$), making it impossible
to generate systems that are fully consistent with the radial velocity
data by independently sampling orbital parameters from the inferred underlying
distributions.

We conclude that the nominal,
edge-on, coplanar two-planet model of Table 2 is fully consistent with the 
current set of radial velocity measurements of the star.
If the actual stellar jitter for GJ 876 is smaller than $6 \,{\rm m\, s^{-1}}$,
then one can contemplate extracting additional information (related, say,
to the inclinations and nodes of planets ``b'' and ``c'', or to additional 
bodies) from the lists of 
radial velocities in Table 1 and in Marcy et al. (2001). 

The three-body fit to the radial velocities indicates that 
the two major planets in the GJ 876 are locked in 
a symmetric configuration, with the resonant arguments
$\theta_{1}=\lambda_c-2\lambda_b+\varpi_c$, and 
$\theta_{2}=\lambda_c-2\lambda_b+\varpi_b$ both librating about zero
degrees. The orbital configuration of the best-fit edge-on coplanar
model of the GJ 876 data set is shown in Figure 1. In this figure, the
positions of the planets are plotted as filled circles at 60 successive
one-half-day intervals beginning on JD 2449710, when the planets were both
near periastron. The positions are plotted in the frame centered
on the star. Also plotted (as clouds of dots) are the positions of the
planets at every one-half-day interval since the epoch of the first Lick 
Observatory data point taken on JD 2449679.6316. The figure shows that the
orbits of the planets do not close, while examination of the time-dependent
osculating orbital elements shows that the periapses of the planets are
precessing at a rate of $\dot{\varpi}=-41^{\circ}\, {\rm yr^{-1}}$. 
This rapid precession is the
primary reason why Keplerian fits to the data show higher $\sqrt{\chi^2}$ values
than the self-consistent three-body fits.

Figure 2 shows the fitted reflex velocity of the star in comparison with
the radial velocity data. The most striking feature of this figure (aside from the
dominant $\sim60 \, {\rm d}$ periodicity) is the modulation arising from
the 8.7 year precession period for the planets' joint line of apsides.
This precession has now been observed for more than one full period, and 
the planets have completed a full librational cycle for both the secular
$\vert \varpi_c - \varpi_b \vert$ resonance argument, as well as the 2:1
resonance arguments $\theta_1$ and $\theta_2$. These librations are manifest 
in the slightly non-sinusoidal envelope of the overall stellar reflex velocity.
The non-Keplerian aspect of the motion is also evident in the wave of
small-amplitude velocity reversals running through the radial velocity 
curve. In the summed Keplerian model, this wave has an asymmetric shape, and
is produced (along with the overall modulation) by the inner planet ``c"
having a fixed period $P_c=30.12\, {\rm d}$ that is slightly less than half
the $P_b=61.02\, {\rm d}$ period of the outer planet.
In the self-consistent fit, the small velocity reversals display a symmetric waveform, and arise
largely from the precession of the inner eccentric orbit and the
librations about the three resonances. For additional related discussion
of the manifestation of the orbital dynamics in the radial velocity curve,
 see Nauenberg (2002)

The primary assumption underlying the fit given in Tables 2 and 3 is that the planetary
orbits are co-planar and are being viewed edge-on. While there is no a-priori
observational evidence to indicate that the system is co-planar, it is likely
that the planets arose from a relatively flat protoplanetary accretion disk.
Numerical integrations of the differential migration of the system which
assume this scenario show that the eccentricity must in general
be forced to higher values than observed before significant mutual inclination
is excited (see also Thommes \& Lissauer 2003). Hence it makes dynamical
sense to prefer co-planar models. We also note that astrometric evidence 
obtained by Benedict et al. (2002) suggests that the outer planet in the
system is being viewed fairly close to an edge-on configuration.

If we assume co-planar inclinations with $i_s<90^{\circ}$ 
and construct a succession of fits, we obtain the run of best-fit
$\sqrt{\chi^{2}}$ values shown by the thick dashed line of Figure 3. 
This sequence
shows a very slight decline in the value of $\sqrt{\chi^{2}}$ as the 
system is tilted from $i_s=90^{\circ}$ ($\sqrt{\chi^{2}}=1.535$)
to $i_s=59^{\circ}$ ($\sqrt{\chi^{2}}=1.525$). For co-planar
inclinations having $i_s<38^{\circ}$, however, 
$\sqrt{\chi^{2}}$ experiences a rapid rise. Similar behavior in the 
$\sqrt{\chi^2}$ profile was observed by both Laughlin \& Chambers (2001)
and Rivera \& Lissauer (2001), although with more radial velocity data, the dip in $\sqrt{\chi^2}$ 
has grown shallower.  Laughlin \& Chambers (2001), and Rivera \& Lissauer (2001)
both intepreted the configuration with the minimum $\sqrt{\chi^{2}}$ as
representing the likely coplanar inclination of the system, whereas Nauenberg (2002)
suggested that the improvement found by those authors in going from $i_s=90^{\circ}$
to $i_s\sim 45^{\circ}$ was not significant.
Our primary aim is thus to ascertain what
significance can be ascribed to this trend in $\sqrt{\chi^{2}}$. That is,
which co-planar inclinations can be ruled out by the dynamical fits
to the data?

In Figure 4, we plot the best-fit osculating eccentricities, $e_b$ and $e_c$ 
as a function of co-planar inclination $90-i_s$. As $i_s$ decreases from $90^{\circ}$,
the fitted planetary masses increase
by $\sin(i)^{-1}$, and the fitted eccentricities also increase. The 
inner planet eccentricity, for example, increases from $e_c=0.22$ at  
$i_s=90^{\circ}$ to $e_c=0.38$ at $i_s=20^{\circ}$. The best fit $\sqrt{\chi^{2}}$
value, however, changes very little in the face of this large eccentricity
increase. This
behavior occurs because the primary non-Keplerian interaction between
the planets is the $\dot{\varpi}=-41^{\circ} {\rm yr}^{-1}$ 
precession (see e.g. Ford 2003). As the masses
of the planets increase, the precession rate also increases. This 
increase, however, can be essentially exactly offset by an increase in
the orbital eccentricities, which act to decrease $\dot{\varpi}$. 

For each of the 100 Monte-Carlo realizations of synthetic
data sets which were previously generated for the
edge-on coplanar system listed in Tables 2 and 3, we perform the 
same procedure of incrementing
the co-planar inclination and obtaining fits. The results are
shown as the cloud of dots in Figure 3, in which thirteen randomly
selected sequences are also plotted as dark lines in order to give
a representative idea of the trends for particular realizations.
These fits show that 
the shallow minimum observed near $i\sim60^{\circ}$ in the fits
to the actual data cannot be believed.

Figures 5, 6 and 7 show the fitted values of the 2:1 and secular libration
widths ($\vert \theta_1 \vert_{max}$, $\vert \theta_2 \vert_{max}$, 
and $\vert \varpi_b-\varpi_c \vert_{max}$)
as a function of co-planar inclination for the Monte-Carlo
realizations. The fits to the actual data (heavy dashed lines) are fully
consistent with the behavior observed in the Monte-Carlo realizations,
providing further evidence that the inclination of the co-planar system
cannot be confidently extracted from the data (assuming Gaussian
stellar jitter with $\sigma=6\,{\rm m\,s^{-1}}$).

The libration width figures indicate that
as the masses of the planets are increased (i.e. as $90-i_s$ increases) the libration
widths $\vert \theta_1 \vert_{max}$, $\vert \theta_2 \vert_{max}$,
and $\vert \varpi_b-\varpi_c \vert_{max}$  all show a decrease,
reaching minimum values near $i_s \sim 45^{\circ}$. This phenomenon occurs
because the librations are more readily sensed in an radial velocity data set for planets of
larger mass. Hence, a given observed perturbation must arise from a smaller 
libration if the planet masses are increased. The increase in $\sqrt{\chi^{2}}$ 
observed for systems with $i_s<30^{\circ}$ is associated with the inability to match
the observed perturbations with further decreases in the libration widths of the
resonances. We note that simulations of resonant capture (Kley et al 2003) and
differential migration favor narrow libration widths. These scenarios would therefore
favor the 
prediction that the system will eventually be found to lie 
in the neighborhood of $i_s\sim45^{\circ}$.

\section{The Prospects for Observing GJ 876 ``c'' in Transit} 

The {\it a priori}  probability that a planet on a Keplerian orbit 
transits its parent star as seen from the line of
sight to Earth is given by,
$${\cal P}_{\rm transit}=0.0045 \left({ {1 {\rm AU}} \over {a}}\right)\left({R_{\star}+R_{\rm pl}\over{R_{\odot}}}
\right)\left({1+e \cos(\pi/2-\varpi) \over{1-e^{2}}}\right) \,\eqno(1)  $$
where $a$ is the semi-major axis of the orbit, $R_{\star}$ and $R_{\rm pl}$ are 
the radii of the star and planet, respectively, $e$
is the orbital eccentricity, and $\varpi$ is the argument of 
periastron referenced to the intersection of the plane
of the sky with the orbital plane, namely the line of nodes. For the inner planet
of the GJ 876 system, this probability is only $\sim1\%$, if we assume
a stellar radius of $R=0.3 R_{\odot}$.
Planet-planet interactions in the GJ 876 system, however, allow the
nodal line of the
inner, less massive planet to precess into transit for a significantly 
wider variety of observationally consistent
non-coplanar configurations. Transits of GJ 876 by the inner planet ``c", if they occur, are
therefore likely to be visible for a period of order two years as the node
of the planetary orbit sweeps across the face of the star. The scientific
opportunities from such transits 
would be somewhat analogous to the opportunities provided by 
the series of mutual eclipses observed
in the Pluto-Charon system in the 1980s (Binzel 1989).
Such configurations require a mutual inclination
between planets ``b'' and ``c''.
Because a transiting configuration is
relatively easy to observe (the transit depth is expected to be of
order $10\%$, this system makes an interesting photometric target
for small-aperture telescopes (e.g. Seagroves et al. 2003).

Using the planetary evolution models computed by Bodenheimer, Laughlin,
\& Lin (2003), and assuming $i_s=90^{\circ}$,
we estimate that the planetary radius of GJ 876 ``c'' should
be 0.93 $R_{\sc JUP}$ if the planet has a solid core, and 1.03 $R_{\sc JUP}$
if it does not. Insolation-driven atmospheric-interior coupling,
which can lead to an increased radius
(see e.g. Guillot \& Showman 2002), is expected to be negligible
for planet ``c''. For an assumed tidal quality factor $Q=10^{6}$, the 
eccentricity damping timescale is of order 250 Gyr (Goldreich \& Soter 1966),
indicating that the energy generated by interior tidal heating should not
affect the planetary radius.
We estimate that the effective temperature
at the planet's $\tau=1$ surface is 210K, assuming an albedo
$a=0.4$.

Benedict et al. (2002) used the FGS instrument on HST to obtain
a preliminary measurement of the inclination of the outer planet in
the GJ 876 system, obtaining a value $i_b=84 \pm 6^{\circ}$.
In order to illustrate the possibility that the inner planet may periodically
experience transit epochs, 
we assume that the orbital plane of the outer planet is coincident with the line of
sight at JD 2449679.6316 ($i_b=90^{\circ}$). We then choose (1) a
specific value for the osculating inclination of the inner planet at the
epoch of the first radial velocity point, as well as (2) the osculating value of the
difference in nodal longitudes at the first radial velocity epoch. With these
parameters fixed, we then obtain a self-consistent fit to the radial velocity data to
determine all the other orbital parameters. When an acceptable fit is
obtained, we integrate the system forward to check for the occurrence of
(inner planet) transits within the next 100 years.

The results are shown in the lower left hand panel of Figure 8, which
shows the result of 1296 such separate self-consistent fits.
In the figure, the fits are organized by the
choice of osculating starting inclination of the inner planet orbit
($y$-axis of the figure panels), and
by the initial angle between the two ascending nodes ($x$-axis of the figure panels).
The nominal edge-on co-planar system therefore
corresponds to the bottom row of cells. Scenarios where the inner planet was
transiting during the last season of observations (and specifically during
the transit epoch near JD 2453000.57)
have their cells colored black.
Systems that start transiting within 100 years of the last radial velocity observation in
Table 1 are indicated by
dark gray (transits to start very soon) to light gray (transits starting in
the year 2103). Some regions of the diagram
contain systems which were transiting during the past ten years, but
which have by now moved out of alignment. 
On average, over the range of configurations plotted in Figure 8, the line of 
nodes of the inner planet precesses at rates of order $-4^{\circ}\, {\rm yr^{-1}}$.

The lower right panel of Figure 8 maps the distribution of $\sqrt{\chi^{2}}$ values
obtained for the 1296 separate self-consistent fits. The lowest values found
are $\sqrt{\chi^{2}}=1.52$, matching the best-fit co-planar $i_s=90^{\circ}$ model.
All of the models have $\sqrt{\chi^{2}}<1.65$. The Monte Carlo analysis of the previous
section thus indicates that they are acceptable fits to 
the radial velocity data, assuming $\sigma=6\,{\rm m\,s^{-1}}$. 
In the top two panels of Figure 8, we plot the libration widths of the secular
apsidal alignment, and the $\theta_1$ resonance argument for each fit. 
These panels show that, for the range 
of mutual inclinations sampled, the resonant conditions are always fulfilled.

\section{Discussion}

Our analysis continues to show that
the Non-Keplerian interaction between the two planets in the GJ 876
system indicates that the planets are participating in both the 2:1 mean
motion resonances, as well as in the secular apsidal resonance. Radial
velocities accumulated over the last four years show that the libration
widths of all three resonances are narrow, which argues for a dissipative
history of differential migration for the system. 

It is interesting to note, however, that the planet-planet interactions are
in a sense quite subtle, and suffer from a degeneracy which prevents 
simultaneously accurate measurement of the eccentricity of the inner planet
and the overall inclination of the system. Extensive Monte-Carlo simulations
suggest that the eccentricity of the inner planet lies in the range 
$0.2<e_c<0.35$, and that the system has $i_s>20^{\circ}$. This situation is based on
an assumption for the stellar jitter of $6\, {\rm m \, s^{-1}}$. If this assumption turns out to
be conservative, and the actual jitter is less, then it will be possible to
obtain considerably better constraints on the orbital parameters of the system,
and as more radial velocities are obtained, perhaps confirm or rule out the presence of
additional small bodies in this remarkable exoplanetary system.

Plausible and detailed histories for the origin of the resonances in the GJ 876 system
were proposed by Lee \& Peale (2001, 2002). In their scenario, the planets originally
formed in low eccentricity orbits with semi-major axes larger than those currently
observed, and with a larger period ratio than the present-day 2:1 commensurability. The
planets then grew large enough to open gaps in the protoplanetary disk.
Hydrodynamic simulations by Bryden et al. (2000), and Kley (2000) suggest that
a residual ring of disk material between two massive planets is rapidly cleared as
a consequence of repeated spiral shock passages from the protoplanetary wakes. This
clearing process appears to require only several hundred orbits after the planets
have been established. After the ring of gas between the planets has vanished,
the planets will experience differential migration. The spiral wake driven
through the outer disk will exert a negative torque on the outer planet, causing it
to spiral inward. The inner planet will either be pushed outward by a remnant
inner disk, or more likely, will retain a more or less constant semi-major axis. 
The inward-migrating
outer planet then captures the inner planet into a low-order mean motion
resonance (which in the case of GJ 876 was the 2:1) and the planets migrate in together. 
Lee \& Peale (2002)
demonstrated this mode of resonant capture for GJ 876 through the use of torqued
three-body simulations. Additional N-body simulations of the GJ 876 precursor system
were performed by a number of authors 
including Snellgrove, Papaloizou \& Nelson (2001), Murray, Paskowitz
\& Holman (2002),
Nelson \& Papaloizou (2002), and 
Beaug\'e, Ferraz-Mello, \& Michtchenko (2004).
More recently, full hydrodynamical simulations by Papaloizou (2003), and
Kley, Peitz \& Bryden (2004) have also demonstrated capture of GJ 876 ``b'' and ``c'' into 
the observed resonances as a consequence of differential migration driven by disk torques.

Once the planets are migrating in resonance in response to outer disk torques, 
the orbits lose angular momentum and energy
at different rates. In the absence of a dissipative mechanism, this mismatch causes the planetary
eccentricities to increase. Lee \& Peale (2002) introduced an ad-hoc eccentricity damping
term to the migration. In cases where eccentricity damping was not used, they found that the
semi-major axes decreased by only 7\% before the eccentricities were pumped to their
observed nominal values ($e_c=0.22$, and $e_b=0.03$). They therefore 
suggested that either (i) the disk
dissipated before the planets were able to migrate very far, or alternately, that (ii) an
effective mechanism exists for eccentricity damping during resonant migration.

Option (i) appears to require fine-tuning in order to provide an
explanation for
the current state of the GJ 876 system. The GJ 876 red dwarf, with $M=0.3 M_{\odot}$,
is nearly one hundred times less luminous
than the Sun. The inner planet, GJ 876 ``c",
with its surface temperature $T\sim 210 \,K$, is not far inside the location of the 
current snow-line of the GJ 876 system. For GJ 876 ``b'', located at $a=0.2 {\sc AU}$, 
we estimate a temperature
at $\tau=1$ of $T_{b}\sim 160 \, K$, which places it at or beyond the present snowline. 
The stellar evolution models of Baraffe et al. (2002), however, indicate that during
contraction phases between 1 and 10 million years when giant planet formation likely took
place, GJ 876 was more than ten times as luminous as it is now.
The possibility of
nearly {\it in situ} formation for the GJ 876 planets is therefore unlikely, but
not fully out of the question
(see e.g. the accretion models of Bodenheimer, Hubickyj, \& Lissauer 2000). Certainly,
the comparatively luminous early phases of M star evolution pose interesting tests for
theories of planet formation.

Option (ii) may also be problematic. Recent 2D hydrodynamical simulations,
such as those of Kley, Peitz, \& Bryden (2004) are able to follow the
planet-planet-disk evolution over secular timescales  $t>5\times10^{4} \, {\rm yr}$
These simulations self-consistently model both the resonance capture
and differential migration processes, and show that eccentricity damping
arising from the disk gas is much smaller than that required by Lee \& Peale (2002)
to explain the current state of the GJ 876 system as arising from significant
differential migration.
Kley, Peitz, \& Bryden (2004) remark, however, that it
remains to be seen whether 3-D hydrodynamic calculations, which incorporate
a more realistic equation of state, and which adequately resolve the gas
flow close to the planets, will provide the needed increase in the eccentricity
damping rate.

The low expected temperature of GJ 876 ``c" leads naturally to speculation that 
a potentially habitable terrestrial world might exist in the system. The usual definition
of the planetary habitability zone, as given in Kasting et al. (1993), 
combined with the stellar properties of GJ 876, suggests that
the habitable zone of GJ 876 is located interior to the orbit of planet ``c'' ($a_c=0.13 {\sc AU}$)
at a radius $r_h\sim0.1 {\sc AU}$. Menou \& Tabachnik (2003) report that terrestrial planets
placed in habitable circular orbits with $0.1 \, {\sc AU} < a < 0.2 \, {\sc AU}$ are rapidly ejected by the outer
two planets. We have verified this conclusion using the updated orbital parameters given
in Table 2.

We remark, however, that the clear history of resonant capture and inward dynamical migration in this system
suggests that a terrestrial-mass object orbiting interior to the two gas giant planets
may have been captured into a 2:1 resonant orbit
with GJ 876 ``c", leading to a high-eccentricity analog of the Laplacian resonant condition
observed among Io, Europa and Ganymede. Such an object would have an orbital period of
order $P\sim15$ days, and a semi-major axis of $a_t=0.08 {\sc AU}$. Numerical experiments
show that stable systems of this sort are readily found in which the resonant argument $\theta_2$
between the planet ``c" 
and the putative interior terrestrial planet is librating, and where the eccentricity
of the terrestrial planet is $e_t \sim 0.3$. If such a system is not fully co-planar, then
one can expect precession of the nodal line, and hence periodically recurring transits.
An Earth-size planet transiting GJ 876 would produce a transit depth of ~0.3\%, which is readily
detectable with modest-aperture telescopes from the ground (Henry 1999).
A habitable planet in the GJ 876 system would display a maxium separation from the
primary star of approximately 20 mas, which places the system within the top 300
candidates among the 1139 nearby stars currently being considered as potential targets for
NASA's TPF mission. \footnote{see the TPF target list at 
http://planetquest.jpl.nasa.gov/Navigator/library/basdtp.pdf}

It is likely that the GJ 876 system will reveal further surprises as it is studied
photometrically and spectroscopically from the ground and from space. Furthermore, even the
present radial velocity data set may harbor much additional information if the
stellar jitter turns out to be smaller than $\sigma=6\,{\rm m\,s^{-1}}$ that we have
assumed in this study.

We thank Drs. Peter Bodenheimer, Eric Ford, Man Hoi Lee, Jack Lissauer, Mike Nauenberg and Eugenio Rivera
for useful discussions.  
This material is based upon work supported by the National Aeronautics and Space Administration
under Grant NNG-04G191G issued through the Terrestrial Planet Finder Foundation 
Science Mission (to GL).
We acknowledge support by
NSF grant AST-0307493 (to SSV), NSF grant AST-9988087 and travel
support from the Carnegie Institution of Washington (to RPB), NASA
grant NAG5-8299 and NSF grant AST95-20443 (to GWM), and by Sun
Microsystems. This research has made use of the Simbad database, 
operated at CDS, Strasbourg, France.

\clearpage

\begin{figure}
%\plotone{ms.fig1.ps}
\caption{
Orbital motion arising from  the 2-planet co-planar dynamical fit to the GJ 876
system listed in Tables 2 and 3. The clouds of small black dots plot the
positions of the planets at every one-half-day interval from JD 2449680 to
JD 2453000, illustrating the range of planetary motion produced by the
precession of the line of apsides.
The connected filled circles plot the positions of the planets at
120 one-half-day intervals beginning on JD 2449710. The two solid lines radiating
from the central star mark the osculating longitudes of periastron, $\varpi_b=149.1^{\circ}$
and $\varpi_c=154.4^{\circ}$ for the planets at JD 2449710. The longitudes $\varpi_b$
and $\varpi_c$
oscillate about alignment with a libration amplitude ${\vert \varpi_c -
\varpi_b \vert}_{\rm max}=34^{\circ}$, and the line of apsides precesses
at a rate $\dot{\varpi}=-41^{\circ} {\rm yr}^{-1}$. The sense of the orbital motion is
counterclockwise as viewed from above.
}
\end{figure}
\clearpage

\begin{figure}
%\plotone{ms.fig2.ps}
\caption{
{\it Top Panel}: Stellar reflex velocity from a 
self-consistent, co-planar, $i_s=90^{\circ}$ three-body integration
compared to the GJ 876 radial velocities. The fit parameters and initial conditions
for the integration are listed in Tables 2 and 3. Velocities obtained
at 
Lick Observatory (listed in Marcy et al. 2001) and the velocities
taken at Keck Observatory (listed in Table 1) are shown as
small solid circles. The plotted Lick velocities include a fitted offset between the
telescopes which resulted in $o_2=44.476 {\rm m\,s^{-1}}$ being added to
each of the 16 Lick Observatory measurements.
{\it Bottom Panel}: Residuals to the orbital fit.
}
\end{figure}
\clearpage

\begin{figure}
%\plotone{ms.fig3.ps}
\caption{
$\sqrt{\chi^2}$ values obtained from 
three-body fits to the GJ 876 radial
velocity data as a function of co-planar inclination, 
$90-i$ (heavy dashed line).
Also shown are the $\sqrt{\chi^2}$ values obtained (as a function of 
assumed co-planar inclination) from
fits to Monte-Carlo realizations
of the edge-on configuration listed in Table 2
(black lines and cloud of black dots).
}
\end{figure}
\clearpage

\begin{figure}
%\plotone{ms.fig4.ps}
\caption{
Eccentricity of the inner planet, $e_c$, (connected open symbols)
and the outer planet, $e_b$, (connected filled symbols)
vs. $\sin(i)$ for co-planar 2-planet fits to the GJ 876 radial
velocity data set.
}
\end{figure}
\clearpage

\begin{figure}
%\plotone{ms.fig5.ps}
\caption{
Maximum libration angle $\vert \varpi_c-\varpi_b \vert_{max}$ observed in 
fits to the GJ 876 radial
velocity data (heavy dashed line), along with fits to Monte-Carlo realizations
of the edge-on configuration listed in Table 2
(black lines and cloud of black dots).
}
\end{figure}
\clearpage

\begin{figure}
%\plotone{ms.fig6.ps}
\caption{
Maximum libration of the 2:1 resonant argument $\vert \theta_1 \vert_{max}$ observed in 
fits to the GJ 876 radial
velocity data (heavy dashed line), along with fits to Monte-Carlo realizations
of the edge-on configuration listed in Table 2
(black lines and cloud of black dots).
}
\end{figure}
\clearpage

\begin{figure}
%\plotone{ms.fig7.ps}
\caption{
Maximum libration of the 2:1 resonant argument $\vert \theta_2 \vert_{max}$ observed in 
fits to the GJ 876 radial
velocity data (heavy dashed line), along with fits to Monte-Carlo realizations
of the edge-on configuration listed in Table 2
(black lines and cloud of black dots).
}
\end{figure}
\clearpage

\begin{figure}
%\plotone{ms.fig8.ps}
\caption{
{\it Upper left panel}: 
${\vert \varpi_c - \varpi_b \vert}_{max}$ for fits in which the
2 planets are assumed to be mutually inclined. For all fits, $i_b=90^{\circ}$
at epoch JD 2449679.6316. Fits are gridded according to $\Omega_c-\Omega_b$
($x$-axis of each panel) and $90^{\circ}-i_c$ ($y$-axis of each panel).
The grid cell corresponding to each fit
is color coded and can vary from white 
(${\vert \varpi_c-\varpi_b \vert}_{max} \ge 60^{\circ}$)
to dark
(${\vert \varpi_c-\varpi_b \vert}_{max} \le 10^{\circ}$).
{\it Upper right panel}: same as upper left panel, except ${\theta_1}_{max}$ is
plotted with color coding ranging from white
(${\theta_1}_{max} \ge 20^{\circ}$)
to dark
(${\theta_1}_{max} \le 5^{\circ}$).
Lower right panel: same as upper left panel, except $\sqrt{\chi^{2}}$ is
plotted for each fit with color coding ranging from white
($\sqrt{\chi^2} \ge 1.65$) 
to dark
($\sqrt{\chi^2} \le 1.52$).
Lower left panel: same as upper left panel, except the starting epochs for
transits of planet ``c" are plotted
for each fit with color coding ranging from
light gray (transits in year 2100),
to dark (transiting during the first line of sight passage after epoch
JD 2452988.724 of the last radial velocity measurement in Table 1).
}
\end{figure}
\clearpage

\begin{deluxetable}{rrr}
\tablenum{1}
\tablecaption{Measured Velocities for GJ 876 (Keck)}
\label{measuredvels}
\tablewidth{0pt}
\tablehead{
JD$~~$ & RV$~~$ & Unc.$~~$ \\
(-2440000)   &  (m s$^{-1}$) & (m s$^{-1}$)
}
\startdata
\tableline
10602.093  &   275.000  &   4.83744 \\
10603.108  &   293.541  &   4.87634 \\
10604.118  &   283.094  &   4.67169 \\
10605.110  &   280.726  &   5.52761 \\
10606.111  &   263.544  &   4.98461 \\
10607.085  &   233.736  &   4.69071 \\
10609.116  &   150.489  &   5.45623 \\
10666.050  &   280.291  &   5.21384 \\
10690.007  &  -166.391  &   5.10688 \\ 
10715.965  &   143.299  &   4.61302 \\
10785.704  &   311.515  &   8.24044 \\
10983.046  &  -105.733  &   4.95257 \\
10984.094  &  -123.184  &   5.06684 \\
11010.045  &  -94.0837  &   4.60856 \\
11011.102  &  -73.0974  &   3.64249 \\
11011.986  &  -45.1522  &   2.97916 \\
11013.089  &  -18.1096  &   5.04988 \\
11013.965  &   1.82459  &   3.41393 \\
11043.020  &  -88.9192  &   4.74611 \\
11044.000  &  -115.336  &   4.14800 \\ 
11050.928  &  -159.471  &   4.70908 \\
11052.003  &  -144.716  &   5.26416 \\
11068.877  &  -132.122  &   4.85189 \\
11069.984  &  -103.148  &   4.23548 \\
11070.966  &  -109.364  &   4.57718 \\ 
11071.878  &  -78.1619  &   4.73338 \\
11072.938  &  -62.7263  &   4.78583 \\
11170.704  &  -125.859  &   6.25166 \\
11171.692  &  -134.732  &   6.13061 \\
11172.703  &  -114.607  &   5.41503 \\
11173.701  &  -110.987  &   6.15440 \\
11312.127  &  -145.816  &   4.74986 \\
11313.117  &  -147.122  &   5.23696 \\
11343.041  &   30.2319  &   4.84609 \\
11368.001  &  -194.527  &   4.32588 \\
11369.002  &  -198.763  &   4.71886 \\
11370.060  &  -178.623  &   4.44845 \\
11372.059  &  -175.318  &   8.09112 \\
11409.987  &  -92.8259  &   4.43375 \\
11410.949  &  -92.9346  &   4.29902 \\
11411.922  &  -105.284  &   4.83919 \\
11438.802  &  -72.3787  &   4.40473 \\
11543.702  &  -155.604  &   6.97488 \\
11550.702  &  -195.472  &   6.43277 \\
11704.103  &   107.247  &   4.76474 \\
11706.108  &   60.9448  &   5.38786 \\
11755.980  &   251.808  &   7.45259 \\
11757.038  &   233.575  &   5.95792 \\
11792.822  &  -220.933  &   4.59942 \\
11883.725  &   171.247  &   5.53578 \\
11897.682  &   39.5285  &   6.05218 \\
11898.706  &   37.9805  &   5.67651 \\
11899.724  &   27.5835  &   6.14660 \\
11900.704  &   11.2978  &   5.16144 \\
12063.099  &   197.931  &   5.85069 \\
12095.024  &  -242.481  &   5.64944 \\
12098.051  &  -281.799  &   5.68766 \\
12099.095  &  -267.919  &   5.08161 \\
12100.066  &  -275.558  &   5.42508 \\
12101.991  &  -254.637  &   5.08017 \\
12128.915  &   122.199  &   6.13512 \\
12133.018  &   55.8623  &   5.23130 \\
12133.882  &   59.7818  &   5.75232 \\
12160.896  &  -256.467  &   5.11438 \\
12161.862  &  -269.742  &   5.63196 \\
12162.880  &  -237.342  &   5.50410 \\
12188.909  &   95.7486  &   5.98643 \\
12189.808  &   99.2101  &   6.39510 \\
12236.694  &   164.781  &   6.22869 \\
12238.696  &   187.889  &   5.50345 \\
12242.713  &   197.089  &   6.73589 \\
12446.071  &   75.3063  &   6.21285 \\
12486.917  &   185.162  &   3.98475 \\
12487.124  &   174.897  &   3.63466 \\
12487.919  &   171.914  &   4.32865 \\
12488.127  &   170.714  &   3.83680 \\
12488.945  &   149.798  &   2.28548 \\
12514.867  &  -129.741  &   5.24131 \\
12515.873  &  -156.261  &   4.97779 \\
12535.774  &   32.8722  &   5.56979 \\
12536.024  &   41.7095  &   5.54259 \\
12536.804  &   74.7051  &   6.10848 \\
12537.013  &   66.7469  &   4.94657 \\
12537.812  &   76.0011  &   5.38256 \\
12538.014  &   83.6270  &   5.03001 \\
12538.801  &   107.662  &   4.73423 \\
12539.921  &   123.450  &   5.36501 \\
12572.713  &  -43.1787  &   4.71435 \\
12572.919  &  -53.0618  &   5.07898 \\
12573.742  &  -66.6653  &   4.44116 \\
12573.878  &  -73.8528  &   4.31492 \\
12574.763  &  -112.285  &   4.29373 \\
12574.940  &  -116.120  &   4.66948 \\
12575.719  &  -136.679  &   4.42299 \\
12600.751  &   118.311  &   3.86951 \\
12601.750  &   125.363  &   3.89455 \\
12602.721  &   147.247  &   4.25502 \\
12651.718  &  -129.194  &   8.13000 \\
12807.028  &   148.787  &   5.36857 \\
12829.008  &  -254.556  &   4.39146 \\
12832.080  &  -180.797  &   4.86352 \\
12833.963  &  -135.235  &   4.79979 \\
12835.085  &  -100.468  &   4.85671 \\
12848.999  &   141.070  &   6.62912 \\
12850.001  &   127.450  &   6.13066 \\
12851.057  &   121.834  &   5.86209 \\
12854.007  &   84.0791  &   5.10877 \\
12856.016  &   112.441  &   5.21600 \\
12897.826  &  -55.1842  &   4.93230 \\
12898.815  &  -26.9680  &   4.83134 \\
12924.795  &   215.024  &   5.67363 \\
12987.716  &   198.162  &   7.74271 \\
12988.724  &   194.946  &   5.98031 \\
\enddata
\end{deluxetable}

\newpage

\begin{deluxetable}{lll}
\tablenum{2}
\tablecaption{Co-Planar Fit to GJ 876 Radial Velocity Data}
\label{orbitalels}
\tablewidth{0pt}
\tablehead{
Parameter$~~$ & Planet c$~~$ & Planet b$~~$ \\
}
\startdata
\tableline
$P$ (d)              & 30.38 $\pm$ 0.03 & 60.93 $\pm$ 0.03      \\
$M$                 & $0 \pm  15^{\circ}$  & $186 \pm 13^{\circ}$  \\
$e$                  & 0.218 $\pm$ 0.002 & 0.029 $\pm$ 0.005  \\
$i$ fixed            & $90.0^{\circ}$ & $90.0^{\circ}$  \\
$\varpi$             & $154.4 \pm 2.9^{\circ}$ & $149.1 \pm 13.4^{\circ}$ \\
$m$ & 0.597 $\pm$ 0.008 $M_{Jup}$& 1.90 $\pm$ 0.01 $M_{Jup}$\\
$o_1$ ${\rm m \, s^{-1}}$ & -8.732 &  \\
$o_2$ ${\rm m \, s^{-1}}$ & 44.476 &  \\
transit epoch        & JD 2453000.57 $\pm$ 0.22 &  \\
${\vert\varpi_c-\varpi_b\vert}_{max}$ & $34 \pm 11^{\circ}$ & \\
${\theta_1}_{max}$    & $7.0 \pm 1.8^{\circ}$ & \\
${\theta_2}_{max}$    & $34 \pm 12^{\circ}$ & \\
epoch & JD 2449679.6316 & \\
\enddata
\end{deluxetable}

\newpage
\begin{deluxetable}{llll}
\tablenum{3}
\tablecaption{Cartesian Initial Conditions for Co-Planar Fit to GJ 876 Radial Velocity Data}
\label{cartesians}
\tablewidth{0pt}
\tablehead{
Parameter$~~$ & Star$~~$ & Planet c$~~$ & Planet b$~~$ \\
}
\startdata
\tableline
Mass (gm)  & $6.36515181\times10^{32}$ & $1.13341374\times10^{30}$ & $3.59700414\times10^{30}$ \\
$x\,\, {\rm cm}$  & $0.0$ & $-1.3739370\times10^{12}$ & $2.89833447\times10^{12}$ \\
$y\,\, {\rm cm}$  & $0.0$ & $ 6.6185776\times10^{11}$ & $-1.3485766\times10^{12}$ \\
$z\,\, {\rm cm}$  & $0.0$ & $0.0$ & $0.0$ \\
$v_x \,\,{\rm cm \, s^{-1}}$ & $-3.97415664\times10^{3}$ & $-2.53217478\times10^{6}$ & $1.50114165\times10^{6}$ \\
$v_y \,\,{\rm cm \, s^{-1}}$ & $-9.01247643\times10^{3}$ & $-5.26220995\times10^{6}$ & $3.25294014\times10^{6}$ \\
$v_z \,\,{\rm cm \, s^{-1}}$ & $0.0$ & $0.0$ & $0.0$ \\
\enddata
\end{deluxetable}

\end{document}